# High and uniform coherence creation in Doppler broadened double Λ-like atomic system by a train of femtosecond optical pulses

Amarendra K. Sarma* and Pawan Kumar
Department of Physics, Indian Institute of Technology Guwahati, Guwahati-781039, Assam, India.
*Electronic address: aksarma@iitg.ernet.in

We present a detailed analysis of coherence creation in a four level double Λ-like atomic system using a train of ultra-short few-cycle Gaussian pulses. The effect of the Doppler broadening has been analyzed. It is possible to create high coherence across all the velocity groups in the atomic ensemble using pulses with low repetition frequency. The accumulation of coherence at different radial locations within the atomic beam cross-section as a function of the number of pulses in the train is also reported. We find that pulse train with lower repetition frequencies are able to generate high and nearly uniform coherence in a substantial part of the atomic system.

I. INTRODUCTION

Since the first experimental demonstration of coherent population trapping (CPT) in Λ-type system [1], numerous investigations have been made into coherence-induced phenomena like electromagnetically induced transparency (EIT), absorption (EIA), etc. [2-7]. In fact, recently a theoretical investigation shows that CPT can occur even in buckminsterfullerene $C_{60}$ [8]. Arguably, one of the primary reasons behind such an interest is the possibility of modifying the optical properties of the media through quantum interference. Extremely narrow absorption and emission spectral features [9] are associated with CPT and have been identified for use in the implementation of atomic frequency standards [10-11]. Again, the phenomenon of CPT is used to reduce the group velocity of light significantly in certain media [12]. The coherence accumulated in a system can give rise to effects like enhancement of refractive index with zero absorption [13-15], light amplification without population inversion [16], high-harmonic generation [17], etc. Kocharovskaya and Khanin predicted the possibility of using a train of pulses to create coherence in 1986 [18] and later demonstrated experimentally by making use of a comb of optical pulses produced from a mode-locked diode laser [19]. With the recent progress in generation of well-characterized short optical pulses it has become possible to accumulate the coherence in a robust and controllable fashion [20-21]. More recently authors have made theoretical investigations into the generation of coherence using a series of short pulses [22-24]. In particular, Aumiler [20] investigated EIT of a train of femtosecond pulses interacting with the four level atomic systems in the conditions when the pulse repetition period is shorter than the characteristic atomic relaxation times. He considered a special case where the pulse repetition frequency is a subharmonic of the hyperfine splitting of the ground state. Aumiler's works illustrate how the judicious choice of the frequency comb parameters can provide means to effectively control the degree of coherence between the ground-state hyperfine levels for selective velocity groups of atoms. In this work we present a detailed analysis of coherence creation between the two lower states of a four level double Λ-like atomic system using a train of ultra-short few-cycle Gaussian pulses. The effect of Doppler broadening on coherence creation process has been analyzed. Unlike Aumiler, we show that the pulses of low repetition frequency are highly efficient in creating coherence in the system. We have also studied the accumulation of coherence at different radial locations within the atomic beam cross-section as a function of the number of pulses in the train. It is worth mentioning that double Λ-like atomic system, as put by J.H. Eberly as the new workhorse for quantum optics [25], is studied in various contexts such as amplification without inversion [26], high contrast Ramsey fringes [27] and highly efficient four wave mixing [28]. The article is organized as follows. In Sec. II we present the density matrix equations describing the interaction of the four-level system with a train of femtosecond pulses. Sec. III contains the simulated results of the interaction process and their analysis, followed by conclusions in Sec. IV.

## II. THE MODEL

The atomic system considered in this work for creation of coherence is the four-level system shown in Fig.1. These levels can easily be accessed in experiments by considering the well-studied $D_1$ transition hyperfine structure in alkali atoms [29]. For example, state $|1\rangle$ and $|2\rangle$ could correspond to F=2 and F=3 hyperfine levels of the ground state $5S_{1/2}$ of Rb-85 respectively, whereas state $|3\rangle$ and $|4\rangle$ in that case would denote the F'=2 and F'=3 levels of $5P_{1/2}$. Here, we take a representative system and allow it to interact with a train of ultra-short few-cycle femtosecond pulses. The physical situation considered here corresponds to that of a beam of atoms in the gaseous state interacting with the co-propagating laser pulses [30]. A series of pulses in the time domain with well-defined repetition period and phases gives rise to a comb in the frequency domain [31-33]. Thus it is possible to excite a number of different transitions in the atomic system with the corresponding coherent coupling radiations using a single source of laser pulses.

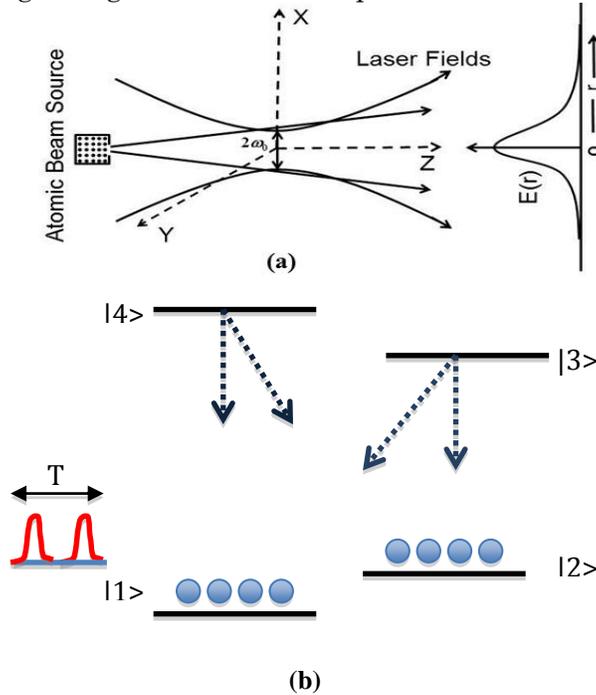

FIG. 1 (a) Schematic diagram of an atomic beam co-propagating with pulsed laser fields (b) Schematic diagram of the energy levels and the coupling radiation

The electric field envelope of the laser pulse-train interacting with the atomic system is described by: $G(r,t) = \sum_{n=0}^{n=N-1} E_0 g(r, t - nT) e^{in\Delta\phi}$, where $g(r,t) = \exp\left(-\left(\left(\frac{r}{W_0}\right)^2 + \left(\frac{t}{\tau_0}\right)^2\right)\right)$ is the Gaussian envelope of a single pulse in the train with the peak amplitude $E_0$ and $\Delta\phi$ is the pulse to pulse phase shift. $W_0$ and $\tau_0$, respectively are the spatial and temporal widths of an individual pulse. $T$ is the pulse repetition period for the train. Within a pulse the electric field profile is given by $E(r,t) = E_0 g(r,t) Cos(\omega_0 t - \Phi(z))$, where $\omega_0$ is the central frequency of the laser pulse. The equation of motion for the density matrix elements, $\rho_{ij}$, describing the time evolution of the considered system is presented in Eq. (1). In this work we have used optical Bloch equations without invoking the rotating wave approximations as we are dealing with intense few-cycle pulse of femtosecond duration. The reasons for using non-rotating wave approximation are well explained by many authors [34-36]. In Eq. (1) we present the optical Bloch equations without invoking the RWA:

$$\frac{d\rho_{11}}{dt} = (\Upsilon_{41}\rho_{44} + \Upsilon_{31}\rho_{33}) + i(\Omega_{13}\rho_{31} - \Omega_{31}\rho_{13}) + i(\Omega_{14}\rho_{41} - \Omega_{41}\rho_{14})$$

$$\frac{d\rho_{22}}{dt} = (\Upsilon_{42}\rho_{44} + \Upsilon_{32}\rho_{33}) + i(\Omega_{23}\rho_{32} - \Omega_{32}\rho_{23}) + i(\Omega_{24}\rho_{42} - \Omega_{42}\rho_{24})$$

$$\frac{d\rho_{33}}{dt} = -(\gamma_{31} + \gamma_{32})\rho_{33} + i(\Omega_{31}\rho_{13} - \Omega_{13}\rho_{31}) + i(\Omega_{32}\rho_{23} - \Omega_{23}\rho_{32})$$

$$\frac{d\rho_{44}}{dt} = -(\gamma_{41} + \gamma_{42})\rho_{44} + i(\Omega_{41}\rho_{14} - \Omega_{14}\rho_{41}) + i(\Omega_{42}\rho_{24} - \Omega_{24}\rho_{42})$$

$$\frac{d\rho_{12}}{dt} = i(\omega_{21}\rho_{12} + \Omega_{13}\rho_{32} + \Omega_{14}\rho_{42} - \Omega_{32}\rho_{13} - \Omega_{42}\rho_{14})$$

$$\frac{d\rho_{13}}{dt} = -(\frac{\gamma_{31}}{2} + \frac{\gamma_{32}}{2})\rho_{13} + i(\omega_{31}\rho_{13} + \Omega_{13}(\rho_{33} - \rho_{11}) + \Omega_{14}\rho_{43} - \Omega_{23}\rho_{12})$$

$$\frac{d\rho_{14}}{dt} = -(\frac{\gamma_{41}}{2} + \frac{\gamma_{42}}{2})\rho_{14} + i(\omega_{41}\rho_{14} + \Omega_{14}(\rho_{44} - \rho_{11}) + \Omega_{13}\rho_{34} - \Omega_{24}\rho_{12})$$

$$\frac{d\rho_{23}}{dt} = -(\frac{\gamma_{31}}{2} + \frac{\gamma_{32}}{2})\rho_{23} + i(\omega_{32}\rho_{23} + \Omega_{23}(\rho_{33} - \rho_{22}) + \Omega_{24}\rho_{43} - \Omega_{13}\rho_{21})$$

$$\frac{d\rho_{24}}{dt} = -(\frac{\gamma_{41}}{2} + \frac{\gamma_{42}}{2})\rho_{24} + i(\omega_{42}\rho_{24} + \Omega_{24}(\rho_{44} - \rho_{22}) + \Omega_{23}\rho_{34} - \Omega_{14}\rho_{21})$$

$$\frac{d\rho_{34}}{dt} = -(\frac{\gamma_{32}}{2} + \frac{\gamma_{31}}{2} + \frac{\gamma_{42}}{2} + \frac{\gamma_{41}}{2})\rho_{34} + i(\omega_{43}\rho_{34} + \Omega_{31}\rho_{14} + \Omega_{32}\rho_{24} - \Omega_{14}\rho_{31} - \Omega_{24}\rho_{32})$$

(1)

In these equations $\gamma_{ij}$, $\omega_{ij} = \omega_i - \omega_j$ and $\Omega_{ij}$ denote the spontaneous decay rate, difference in energy levels and the Rabi frequency corresponding to the pulsed radiation respectively for the energy level pair |i> and |j>. We have assumed that the pair of transitions |1> to |2> and |3> to |4> are dipole forbidden and the rate of decoherence between the two ground states |1> and |2> is negligible and has no influence over the time-scales involved in the interaction with the pulse train. It is very instructive to express the transition frequencies in terms of the repetition frequency, $\nu = 1/T$ of the pulse train and hence we write $\omega_{ij} = 2\pi n_{ij}\nu$ where $n_{ij}$ is a dimensionless number. The Rabi frequencies for various electronic transitions are given by $\Omega_{ij} = \mu_{ij}E(r,t)/\hbar$ where $\mu_{ij}$ is the electric dipole moment for the transition |i> to |j>. We have the relationship $\Omega_{ji} = \Omega_{ij}^*$.

III. RESULTS AND DISCUSSIONS

A. Coherence creation

The dynamics of the double-Λ system, as a result of the interaction with a train of ultra-short optical pulses, exhibits a multitude of features. Depending on the frequencies of electronic transitions available in the atomic system and the relative position of the comb frequencies, the system can be driven to different final states. Fig.2 shows the time evolution of the atomic states under three different conditions of excitation. These results are obtained by numerical integration of Eq. (1) using a fourth order Runge-Kutta scheme. The values of the parameters used in the simulation are: $\Upsilon_{41}=\Upsilon_{42}=\Upsilon_{31}=\Upsilon_{32}= \Gamma/2 =20\times10^6$ s$^{-1}$. This corresponds to a lifetime of 25 ns for the excited states. We also assume that $\Omega_{14}=\Omega_{13}=\Omega_{24}=\Omega_{23}=\Omega$ and that $\Omega_{ij}$'s are all real for simplicity in calculations. Each pulse in the train is characterized by a temporal width given by $\tau_0 = 10 fs$ and spatial width corresponding to $W_0 = 100\ \mu m$. We choose a representative off axis point at $R = 50\ \mu m$ for presenting these results. The on axis peak Rabi frequency is taken to be $\frac{\sqrt{\pi}}{200}\ rad/fs$, which corresponds to a pulse area of $\frac{\pi}{20}$. The pulse-to-pulse phase-shift, Δø, is assumed to be zero in these simulations. In Fig.2 (a) we depict the time evolution of the atomic system for the case when the repetition frequency $\nu = 100$ MHz and the energy levels are given by

$n_{41} = 3.75 \times 10^6$, $n_{21} = 30$ and $n_{43} = 3.6$. The system starts with equal populations in state |1> and |2>, $\rho_{11} = \rho_{22} = 0.5$ and zero initial coherence, $\rho_{12} = 0$. The repetition frequency is such that the condition for two-photon resonance is fulfilled and the system reaches the state with maximum coherence after interaction with about 150 pulses. At this stage, the atoms are in a superposition state of level |1> and |2> given by: $\frac{|1> - |2>}{\sqrt{2}}$. One can also observe that a small amount of population is preferentially transferred into the level |4> whereas $\rho_{33}$ remains close to zero during the entire interaction process. This demonstrates the possibility to isolate the Λ-system formed by the levels |1>, |2> and |4> with very high selectivity even though the frequency spread of a single pulse considered in the train is a few tens of THz.

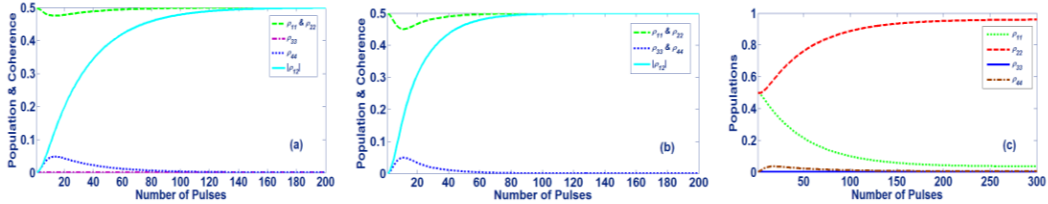

FIG 2. Evolution of populations and coherences under various conditions of excitation (a) ν =100 MHz (b) ν =120 MHz (c) ν = 110.294 MHz.

Fig.2. (b) displays the case in which two photon resonance condition is satisfied with $\nu = 120$ MHz for both the pairs of energy levels (|1>, |2>) and (|3>, |4>) with $n_{21} = 25$ and $n_{43} = 3$. We note that the central frequency of the laser was chosen to be in resonance with the |1> to |4> transition ($\omega_0 = \omega_{41}$). Since the upper levels in both the Λ-systems formed by (|1>, |4> and |2>) and (|1>, |3> and |2>) are in one photon resonance in this excitation scheme, the rate of coherence accumulation is much faster as compared to the case in Fig.2 (a). In Fig.2 (c) we present the result for $\nu = 375/3.4 \approx 110.294$ MHz. This translates into a non-integer value for $n_{21}$ and thus coherence does not develop in the system. On the other hand about 96% of the population is found to occupy the state |2> after interaction with 250 pulses. This is a manifestation of optical pumping in which population is transferred from level |1> to level |2>. These atoms are trapped in state |2> as they cannot decay back to level |1>. It is evident from the results presented in Fig.2 that the repetition frequency of the pulses strongly affects and effectively determines the final state of the atomic system. We now analyze the effect of varying the repetition frequency of the exciting pulses on the coherence accumulation process. Fig.3 (a) depicts the pulse-to-pulse build up of coherence $\rho_{12}$ for three different repetition frequencies. It is important to note that the scheme of excitation considered here as well as the parameter for simulation are the same as in Fig.2. (a).

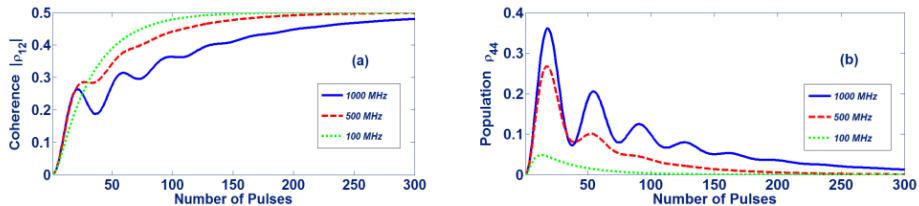

FIG 3. Evolution of (a) coherence $\rho_{12}$ and (b) population in state |4> for different repetition frequencies of the exciting train

We observe that the modulation of $\rho_{12}$ increases with the increase in the repetition frequency and it takes larger number of pulses to achieve the maximum coherence in the system. These modulations are owing to the Rabi oscillations of population between the ground and the excited states [22]. Fig.3 (b) clearly shows that the oscillation of population is much more pronounced in the case of higher repetition frequencies. This behavior can be understood from the fact that the excited state is unable to decay substantially within the short repetition time between two consecutive pulses and it leads to buildup of population in the excited state.

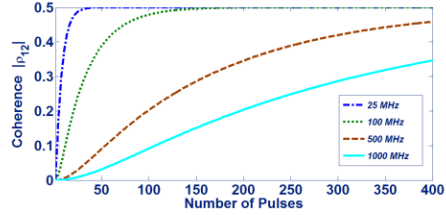

FIG.4. Build-up of coherence $|\rho_{12}|$ in the system with exciting trains of equal average Rabi frequency.

The conditions of excitation in Fig.3 were taken such that the peak power of individual pulse was kept unchanged while its repetition time was varied. We now consider the case in which the average Rabi frequency defined by $\Omega_{avg} = \{\int \Omega(t)\, dt\}/T$ is kept constant. Here the integral is taken over a single pulse of the train. For the simulations we keep $\Omega_{avg}$ unchanged and equal to the case presented in Fig.2. (a) and vary the repetition period $T$ beyond the $T_{Decay}$ where $T_{Decay}$ is the decay lifetime of the excited states. The results obtained from such an interaction are shown in Fig.4. As one would expect, the number of pulses required to accumulate a substantial amount of coherence in the system increases when operating at higher repetition frequencies. Further the modulations in $\rho_{12}$ are absent because only a small amount of population remains in the excited level on average. The result shows that it takes about 35 pulses for the train operating at 25 MHz to drive the system to complete coherence level. It is worthwhile to note that the corresponding repetition period $T = 40\, ns$ is larger than the decay lifetime considered for the excited states. This reflects that exciting pulses with repetition times greater than $T_{Decay}$ remain efficient in accumulating coherence between level $|1\rangle$ and $|2\rangle$.

B. Doppler broadening

A real atomic ensemble at room temperature consists of groups of atoms moving with high velocities. Due to this thermal motion the transition frequencies in the atom are shifted by an amount $\vec{k}.\vec{V}$ in the laboratory frame. Here $\vec{k}$ is the wave vector of the incident radiation and $\vec{V}$ denotes the velocity of the atom. Under the conditions of thermal equilibrium the spread of the velocity distribution and hence the associated Doppler shift, which is Gaussian in nature, can typically be as large as a few hundreds of MHz. This can lead to significant modifications in the response of the system. In this section we present the effects of the Doppler shift on the process of coherence creation in the double $\Lambda$-like atomic systems. We consider a repetition frequency of $\nu = 500$ MHz for the exciting pulses with peak Rabi Frequency of $\frac{\sqrt{\pi}}{200}\, rad/fs$. The central frequency of the laser pulse is taken to be $\omega_0 = \omega_{41} = (2\pi)\, 0.75 \times 10^6 \nu$. This corresponds to a detuning of 1.25 MHz for every $1\, m/s$ speed of the atoms. As we mentioned earlier, the geometry of interaction is assumed such that the atomic beam co-propagates with the pulses and thus the longitudinal velocity of the atoms is responsible for the Doppler shift of the transition frequencies.

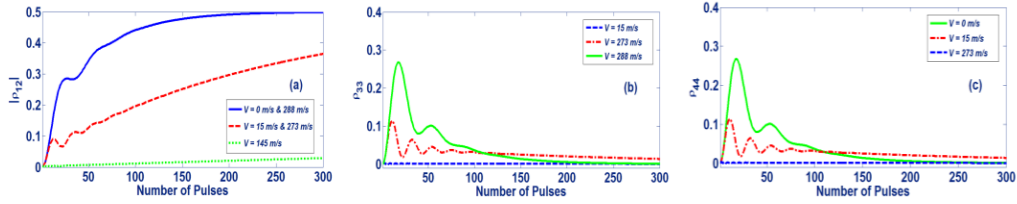

FIG.5. Effect of Doppler shift on (a) coherence creation and evolution of population in (b) state |3> and (c) state |4>.

In Fig.5 (a) we show the build-up of coherence for five different longitudinal velocity groups in the ensemble. It is remarkable that the accumulation process is much faster for the groups with $V = 0\ m/s$ and $288\ m/s$ than the group with $V = 145\ m/s$. This is because whenever the system is in one photon resonance condition the coherence accumulation process is very fast and it takes relatively less number of pulses to drive the system to high coherence. As is evident from Fig.5 (b) and Fig.5 (c), state |3> and state |4> alone are in one photon resonance with the two lower states for the velocity group $V = 288\ m/s$ and $V = 0\ m/s$ respectively. We note that the population in either of the excited states remains very low (~0.1%) for the group with $V = 145\ m/s$. This group is far detuned from both the excited states with respect to one photon resonance condition and thus the coherence creation is quite slow. Fig.6 displays the result of interaction of the femtosecond pulses under the conditions considered in Fig.5 with a broad range of velocity groups in the atomic ensemble, as would be the case for a real atomic system.

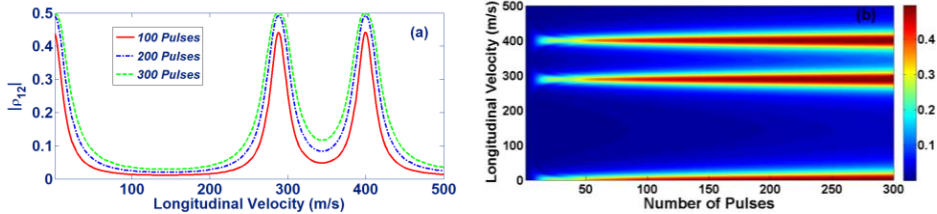

FIG. 6. (a) Variation in coherence induced for different velocity groups (b) Contour plot of variation in $|\rho_{12}|$ due to Doppler effect.

We note that the coherence developed in the atomic system for a given number of excitation pulses varies periodically with the longitudinal velocity of the atoms. The coherence repeats itself after every $400\ m/s$. The behavior stems from the condition that the corresponding Doppler shift and the repetition frequency of the pulse train are equal. Hence in essence the conditions of excitation repeat after every $400\ m/s$ velocity because the separation between two successive teeth of the frequency comb is 500 MHz. We also note that within a period there are two maxima and minima of coherence. As we have explained through Fig.5, the two maxima are a result of the fulfillment of resonance condition involving each of the two excited states. On the other hand we find that when both the upper quantum states are equally and far detuned from the teeth of the frequency comb, the coherence accumulated in the system attains minima. We have seen that Doppler shift of the energy levels can result into significant departures from near complete coherence for specific velocity groups in the atomic ensemble. In fact we have demonstrated that it is possible to achieve high amount of coherence in specific velocity groups by judiciously choosing the repetition frequency of the train and the central frequency of the pulses to manipulate the teeth of the resulting frequency comb.

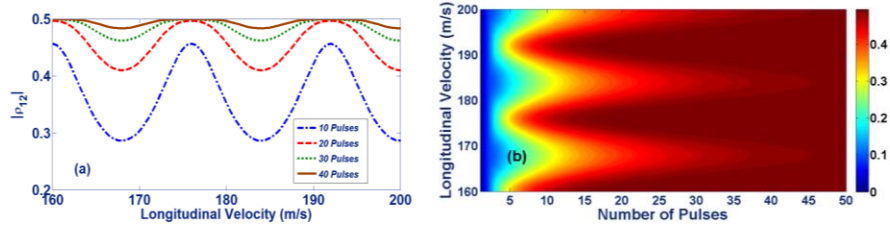

FIG.7. Accumulation of $|\rho_{12}|$ for different velocity groups in the ensemble with the exciting train made up of a few pulses of large pulse areas.

Motivated by the results displayed in Fig.4, we now investigate the possibility of creating high coherence across all the velocity groups in the ensemble. Fig 7 (a) and (b) show the coherence $|\rho_{12}|$ accumulated by a pulse train operating with the repetition frequency of 20 MHz. The average Rabi frequency and other pulse parameters were kept the same as in Fig 4 for these simulations. We find that with just 40 pulses it is possible to accumulate $|\rho_{12}|$ to values more than 97 % in all the velocity groups. It is remarkable that the dip in coherence as it varies periodically with the longitudinal velocity is very small. It is well known that the magnitude of Rabi frequency determines the rate of coherence creation between the ground states of a $\Lambda$-type system with respect to the number of pulses required in the train. It is known that a larger Rabi Frequency leads to complete coherence creation with relatively less number of pulses [22,23]. For the Gaussian pulse profile as considered in this work, amplitude of the electric field and hence the associated Rabi frequency decreases exponentially as one moves away from the common axis of the atomic beam and pulse train. We display in Fig.8 the accumulation of coherence, $|\rho_{12}|$, at different radial locations within the atomic beam cross-section as a function of the number of pulses in the train.

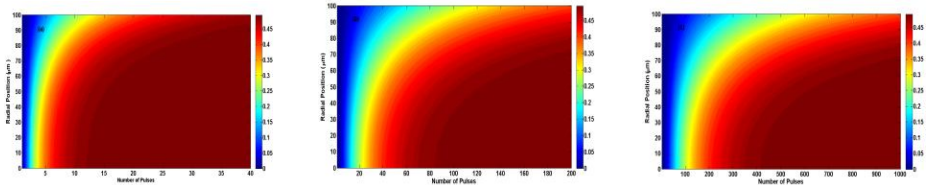

FIG .8. Variation of coherence with radial position with (a) repetition frequency 20 MHz (b) repetition frequency 100 MHz (c) repetition frequency 500 MHz.

It is remarkable that a significant portion of the atomic beam develops near complete coherence when treated with pulses of Gaussian profile. Here, we have presented three cases with repetition frequencies of 20, 100 and 500 MHz. The average Rabi frequencies for these cases have been kept the same as in Fig. 4. We observe that the pulses of low repetition frequency are highly efficient in creating coherence in the system. Fig 8(a) shows that for pulses operating at 20 MHz repetition frequency near-complete coherence is established up to off axis positions as far as 95 µm with just 40 pulses. On the other hand with 200 and 1000 optical pulses operating at 100 and 500 MHz respectively the coherence generated is significantly lower. This result along with that presented in Fig 7 indicate that pulse train with lower repetition frequencies are able to generate high and nearly uniform coherence in a substantial part of the atomic system.

## IV. CONCLUSIONS

In conclusions, we have presented a detailed analysis of coherence creation in a four level double Λ-like atomic system using a train of ultra-short few-cycle Gaussian pulses. The effect of Doppler broadening on coherence creation process has been analyzed. The pulse train with repetition times greater than the decay lifetime of the excited states is efficient in creating coherence between levels |1> and |2>. It is possible to create high coherence across all the velocity groups in the atomic ensemble using pulses with low repetition frequency. The accumulation of coherence at different radial locations within the atomic beam cross-section as a function of the number of pulses in the train is also reported. We find that pulse train with lower repetition frequencies are able to generate high and nearly uniform coherence in a substantial part of the atomic system. Based on this work, similar investigations can be carried out for other types of four-level atomic systems such as ladder-like, Y-type and inverse Y-type atomic systems. The suggested scheme may find practical applications, where a very high and near uniform coherence is required in the entire system. It may be of interest for various applications related to spectrum squeezing, coherent population trapping and transfer, spectral narrowing and elimination, and gain without population inversion, and so on.